\documentstyle[12pt, a4]{article}
\begin{document}

\hfill DFPD 99/Th/32
\vspace{2.0cm}
(August 1999)
\centerline{\Large\bf Finding neutral Higgs bosons }
\centerline{\Large\bf in a 
two-Higgs-doublet model}
\vspace{2.0cm}
\centerline{\Large\bf with spontaneous CP violation}
\centerline{I.Vendramin}
\centerline{\sl Dipartimento di Fisica,Universit\'{a} di Padova,}
\centerline{\sl Istituto Nazionale di Fisica Nucleare,Sezione di 
Padova}
\vspace{3.0cm}
\centerline{e-mail:vendramin@pd.infn.it}
\centerline{\bf Abstract}
We consider a particular two-Higgs-doublet model with spontaneous 
CP violation, where two neutral Higgses have no definite CP properties and 
the third one is CP-even. In 
this model, as the parameter $\theta$ of CP violation increases,
the masses of the neutral Higgs bosons are drawn towards the mass of the 
lightest one and the chance of 
 detection of neutral Higgses
at an $e^+e^-$ collider in $Z+Higgs$,$b\overline{b}+$Higgs and 
$t\overline{t}+$Higgs production increases with respect to the CP 
conserving case.
\vfill\eject
\section{\bf Introduction}
The origin of CP violation is among the most important physics issues.
An elegant way of introducing CP violation is based on the enlargement of 
the Higgs sector of the Standard Model [1][2][3][4].
 
Already the 
simplest extension of the SM, with two scalar Higgs-doublets, predicts 
the existence of five physical Higgs bosons. Among them, two are a 
charged pair and three are electrically neutral.

In the CP conserving version of the two-Higgs-doublet model (2HDM), two 
neutral Higgs bosons are CP-even and one is CP-odd. But, for example in 
the minimal supersymmetric model, with soft supersymmetric CP violating 
phases, the neutral Higgs states will mix beyond the Born approximation, 
leading to CP-impure mass eigenstates [5].

However, in a general 2HDM the Higgs sector itself may also generate 
spontaneous and/or explicit CP violation. Then, in a CP violating 2HDM 
the neutral Higgs states mix already at the tree level; the mass 
eigenstates have indefinited CP parity, i.e., the neutral Higgses have 
both scalar and pseudoscalar Yukawa couplings to quarks and leptons [6].

 In this note, we consider the neutral Higgs production in the higher 
energy $e^+e^-$ colliders, in a particular 2HDM with spontaneous CP 
violation. In this model, let $h_1,h_2,h_3$ to be the three neutral 
Higgs bosons in the order of increasing mass, $h_3$ is CP-even and 
$h_1,h_2$ are CP-impure mass states, depending on the relative phase 
$\theta$ of the vacuum expectation values of the Higgs doublets. In the 
limit of CP conservation, $\theta\to{0}$, $h_1$ becomes CP-odd. 
Meanwhile $h_2$ becomes CP-even and degenerates 
in mass with $h_3$; but their mass (for 
$m(h_1)\ne{0}$) is divergent. In the opposite limit, 
$\theta\to\frac{\pi}{2}$, the CP-impure $h_1$ and $h_2$ degenerate in 
mass and $m(h_3)\to\sqrt{2}m(h_1)$. 

In $e^+e^-$ collisions the processes of production of 
neutral Higgs bosons are (1) the Higgs-strahlung $e^+e^-\to{Zh_k}$, (2) the 
Higgs pair production $e^+e^-\to{h_jh_k}$ and (3) 
the Yukawa processes with Higgs radiation off (heavy) fermions 
$e^+e^-\to{f\overline{f}}\to{f\overline{f}h_k}$. In order to treat these 
processes on the same footing, we consider the $f\overline{f}h_k$ final 
state at future $e^+e^-$ colliders. The process (1) contributes to this 
final state when $Z\to{f\overline{f}}$ and the process (2) when 
$h_j\to{f\overline{f}}$. Now, the unitarity of the 
model implies in general a number of sum rules for the Higgs-gauge boson 
[7][8] and Higgs-fermion [9][10] couplings. (In our particular 2HDM they 
have a very simple form.) As well discussed in ref.(9), these sum rules 
guarantee the detection in $e^+e^-$ collisions of any neutral Higgs boson 
that is sufficiently light to be cinematically accessible in (a) 
Higgs-strahlung {\em and} Higgs pair production or (b) Higgs-strahlung 
{\em and} $b\overline{b}+$Higgs {\em and} $t\overline{t}+$Higgs.

In our particular model, as the parameter $\theta$ of CP 
violation, increases, the masses of the neutral 
Higgs bosons are drawn towards the 
mass of the lightest one. We can have two or even all three neutral 
Higgses "relatively" light. This reduces their production thresholds and 
increases the chance of their finding in the above-mentioned processes. 
In this sense, we are more interested on the cumulative 
$\sum_{1=k}^3{\sigma{(e^+e^-\to{f\overline{f}h_k})}}$ rather than the single 
cross section $\sigma{(e^+e^-\to{f\overline{f}h_1})}$. 

The paper is organized as follows. In section 2, we present our 
two-Higgs-doublet model with spontaneous CP violation and the 
corresponding ZZ-Higgs, Z-Higgs-Higgs and Higgs Yukawa couplings. In 
section 3, we present the cross section formula for 
$e^+e^-\to{f\overline{f}h_k}$ processes. In section 4 we comment on some 
numerical examples of $\sum_k{\sigma{(e^+e^-\to{f\overline{f}h_k})}}$, 
for $f=t,b$.

\section{\bf The two-Higgs-doublet model with spontaneous CP violation}

The 2HDM of the electroweak interaction is the extension of the Standard 
Model by an extra SU(2) Higgs doublet. The two Higgs doublets $\Phi_1$ 
and $\Phi_2$ are assumed to couple to quarks and leptons in such a way 
that there are no flavour-changing neutral coupling at the tree level. 
This natural flavour conservation constraint is enforced (see, for 
instance [11]) customarily by imposing a discrete Z$_2$ symmetry $$
\Phi_2\to{-}\Phi_2\ \ \ \ \ \ \ u_{iR}\to{-u}_{iR}  $$ The requirement 
that the potential breaks this symmetry only softly, excludes terms as 
$(\Phi_1^{\dagger}\Phi_2)[\lambda_6(\Phi_1^{\dagger}\Phi_1)+\lambda_7
(\Phi_2^{\dagger}\Phi_2)]+h.c.$ with operator dimension four, because they 
breaks Z$_2$ symmetry in a hard way. The renormalizable scalar potential 
has the form $$
V(\Phi_1,\Phi_2)=m_1^2\Phi_1^{\dagger}\Phi_1+m_2^2\Phi_2^{\dagger}\Phi_2-
(m_3^2\Phi_1^{\dagger}\Phi_2+h.c.)+\lambda_1(\Phi_1^{\dagger}\Phi_1)^2+
\lambda_2(\Phi_2^{\dagger}\Phi_2)^2+ $$ $$ 
+\lambda_3(\Phi_1^{\dagger}\Phi_1)(\Phi_2^{\dagger}\Phi_2)+\lambda_4
(\Phi_1^{\dagger}\Phi_2)(\Phi_2^{\dagger}\Phi_1)+\frac{\lambda_5}{2}[(
\Phi_1^{\dagger}\Phi_2)^2+h.c.] \eqno(1) $$
We have explicit CP breaking if $\xi=$Im$(m_3^4\lambda_5^*)\ne{0}$. We 
assume $\xi=0$. 

The potential is CP invariant and all parameters are 
real. Minimization of the potential yields the vacuum expectation values 
of the two Higgs fields (the phase of $\Phi_1$ can be adjusted such that 
the v.e.v. of $\Phi_1$ is real and positive, and the phase of $\Phi_2$ 
such that $\lambda_5$ is real and positive) $$  
<\Phi_1>=\frac{v_1}{\sqrt{2}}\ \ \ \ \ \
 \ \ <\Phi_2>=\frac{v_2}{\sqrt{2}}e^{i\theta} \eqno(2) $$ where $v_1,v_2$ 
are real and positive parameters satisfying the experimental constraint 
$v=\sqrt{v_1^2+v_2^2}=$246 Gev. The requirement that our vacuum be at 
least a stationary point of the potential results in the following 
constraints ($\tilde\lambda\equiv\lambda_3+\lambda_4; 
s_{\theta}=\sin\theta; c_{\theta}=\cos\theta$) $$
m_1^2=-\lambda_1v_1^2+\frac{1}{2}(\lambda_5-\tilde\lambda)v_2^2 \eqno(3) 
$$ $$ m_2^2=-\lambda_2v_2^2+\frac{1}{2}(\lambda_5-\tilde\lambda)v_1^2 
\eqno(4) $$ $$ s_{\theta}(m_3^2-\lambda_5v_1v_2c_{\theta})=0 \eqno(5) $$
The last condition entails the interesting case of a non-zero CP 
violating phase $\theta$, provided $|m_3^2/\lambda_5v_1v_2|<1.$ 

After SU(2)$\times$U(1) gauge symmetry breaking, in the neutral Higgs 
sector the would-be Goldstone boson which is eaten in giving mass to 
the Z boson, is the combination 
$\sqrt{2}(c_{\beta}$Im$\phi_1^0+s_{\beta}$Im$\phi_2^0)$ where the mixing 
angle $\beta$ is $\tan\beta=v_2/v_1$. The physical neutral Higgs bosons 
are the eigenstates of the mass$^2$ matrix 
$M^2_{ij}=\frac{v^2}{2}\mu_{ij}$ in the remaining three neutral degrees 
of freedom. The elements of the reduced matrix $\mu_{ij}$, in the basis 
$(\sqrt{2}$Re$\phi_1^0,\sqrt{2}$Re$\phi_2^0,\sqrt{2}\big(s_{\beta}$Im$\phi_1^0-
c_{\beta}$Im$\phi_2^0)\big)\equiv{(\varphi_1^0,\varphi_2^0,\chi_c^0)}$, read 
[12] 
$$\mu_{11}=2\lambda_1c_{\beta}^2+\lambda_5s_{\beta}^2c_{\theta}^2$$
$$\mu_{12}=(\tilde\lambda-\lambda_5s_{\theta}^2)s_{\beta}c_{\beta}$$
$$\mu_{13}=\lambda_5s_{\beta}s_{\theta}c_{\theta}$$
$$\mu_{22}=2\lambda_2s_{\beta}^2+\lambda_5c_{\beta}^2c_{\theta}^2$$
$$\mu_{23}=\lambda_5c_{\beta}s_{\theta}c_{\theta}$$
$$\mu_{33}=\lambda_5s_{\theta}^2 \eqno(6)$$
where we have used (3),(4)and (5) to exclude $m_1,m_2,m_3$ from the 
equations.

Going to the basis 
$(c_x\varphi_1^0-s_x\varphi_2^0,s_x\varphi_1^0+c_x\varphi_2^0,\chi_c^0)$, 
the new matrix elements $$ 
\mu^{\prime}_{13}=-\lambda_5s_{\theta}c_{\theta}s_{x-\beta} \eqno(7) $$ 
$$ 
\mu^{\prime}_{12}=(c_x^2-s_x^2)(\tilde\lambda-\lambda_5)s_{\beta}c_{\beta}+
s_xc_x(2\lambda_1c_{\beta}^2-2\lambda_2s_{\beta}^2)-\lambda_5c_{\theta}^2
s_{x-\beta}c_{x-\beta} \eqno(8) $$ can vanish if $$
x=\beta\ \ \ \ \ \ \ \ \ \ \ \ \ \ (2\lambda_1+
\tilde\lambda-\lambda_5)c_{\beta}^2=
(2\lambda_2+\tilde\lambda-\lambda_5)s_{\beta}^2(\equiv\zeta) \eqno(9) $$ 
In this case $$ \mu^{\prime}_{11}=\zeta+\lambda_5-\tilde\lambda \ \ \ \ \ 
\ \ \ \ \mu^{\prime}_{22}=\zeta+\lambda_5c_{\theta}^2 $$ $$ 
\mu^{\prime}_{33}=\lambda_5s_{\theta}^2\ \ \ \ \ \ \ \ \ 
\mu^{\prime}_{23}=\lambda_5s_{\theta}c_{\theta} \eqno(10) $$
Therefore $$c_{\beta}\varphi_1^0-s_{\beta}\varphi_2^0 \eqno(11) $$ is a mass 
eigenstate with reduced mass$^2$ $\zeta+\lambda_5-\tilde\lambda$. The other 
two eigenstates are $$ 
-s_{\psi}(s_{\beta}\varphi_1^0+c_{\beta}\varphi_2^0)+c_{\psi}\chi_c^0\ \ 
\ \ \ \ \ 
c_{\psi}(s_{\beta}\varphi_1^0+c_{\beta}\varphi_2^0)+s_{\psi}\chi_c^0 
\eqno(12) $$ and have the reduced masses$^2$ 
$\frac{1}{2}(\zeta+\lambda_5-\sqrt{\zeta^2+\lambda_5^2+2\lambda_5\zeta
\cos{2\theta}})$ and 
$\frac{1}{2}(\zeta+\lambda_5+\sqrt{\zeta^2+\lambda_5^2+2\lambda_5\zeta
\cos{2\theta}})$ respectively. In eq.(12) the mixing angle $\psi$ is $$
\sin{2\psi}=\frac{\lambda_5\sin{2\theta}}{\sqrt{\zeta^2+\lambda_5^2+2\lambda_5
\zeta\cos{2\theta}}}; \cos{2\psi}=\frac{\zeta+\lambda_5\cos{2\theta}}
{\sqrt{\zeta^2+\lambda_5^2+2\lambda_5\zeta\cos{2\theta}}} \eqno(13) $$ 
This is the model we assume, with a further condition which makes it 
very simple and attractive from the point of view of the mass spectrum:  
with $$ \zeta=\lambda_5\ \ \ \ \ \ \ \tilde\lambda=0 
\eqno(14) $$ we have $$ \psi=\frac{\theta}{2} \eqno(15) $$ and the 
masses$^2$ become 
$\lambda_5v^2,\lambda_5v^2\sin^2\psi,\lambda_5v^2\cos^2\psi$. Therefore, 
in this model we identify (without loss of generality we choose 
$-\frac{\pi}{2}\le\theta\le{+}\frac{\pi}{2}$, that is 
$-\frac{\pi}{4}\le\psi\le{+}\frac{\pi}{4}$) 

\vspace{0.5cm}
$h_1=\sqrt{2}[-s_{\psi}(s_{\beta}$Re$\phi_1^0+c_{\beta}$Re$\phi_2^0)+c_{\psi}
(s_{\beta}$Im$\phi_1^0-c_{\beta}$Im$\phi_2^0)]$

\vspace{0.5cm}
$h_2=\sqrt{2}[c_{\psi}(s_{\beta}$Re$\phi_1^0+c_{\beta}$Re$\phi_2^0)+s_{\psi}
(s_{\beta}$Im$\phi_1^0-c_{\beta}$Im$\phi_2^0)]$ 

\vspace{0.5cm}
$h_3=\sqrt{2}(c_{\beta}$Re$\phi_1^0-s_{\beta}$Re$\phi_2^0)$ \hfill (16)

\vspace{0.5cm}
 Clearly $h_3$ is CP-even, $h_1$ and $h_2$ generally are 
CP-impure. The masses satisfy a Pythagorical sum rule $$ 
m(h_1)^2+m(h_2)^2=m(h_3)^2 \eqno(17) $$
For given $m(h_1)\ne{0}$, we have $m(h_2)=m(h_1)/|\tan{\psi}|$ and 
$m(h_3)=m(h_1)/|\sin{\psi}|$. If $\theta\to\pm\frac{\pi}{2}$ 
($\psi\to\pm\frac{\pi}{4}$) then $m(h_1)=m(h_2)=m(h_3)/\sqrt{2}$; if 
$\theta\to{0}$ ($\psi\to{0}$) then $h_1$ becomes CP-odd and $h_2$ 
becomes CP even but 
$m(h_2)\to{m(h_3)}\to\infty$.

Once the model is defined, we can give the explicit forms of the 
couplings. The Yukawa interactions of the $h_i$ $(i=1,2,3)$ mass 
eigenstates are given by $$ {\cal L}=\frac{m_u}{v}\overline{u}\big[\frac
{s_{\psi}-ic_{\psi}\gamma_5}{t_{\beta}}h_1-\frac{c_{\psi}+is_{\psi}\gamma_5}
{t_{\beta}}h_2+h_3\big]u+ $$ $$ +\frac{m_d}{v}\overline{d}[(s_{\psi}-ic_{\psi}
\gamma_5)t_{\beta}h_1-(c_{\psi}+is_{\psi}\gamma_5)t_{\beta}h_2-h_3]d  
\eqno(18) $$
for up-type and down-type quarks and similarly for charged leptons.

For large $t_{\beta}$, eq.(18) shows the typical enhancement of the 
couplings to down-type quarks over the coupling to up-type quarks as 
regards $h_1$ and $h_2$, but this is not the case of $h_3$.

Let ${\cal L}=\overline{f}(S_i^f+iP_i^f\gamma_5)fh_i$ and 
$\hat{S}^f_i=\frac{v}{m_f}S_i^f$, $\hat{P}^f_i=\frac{v}{m_f}P_i^f$, the 
check of the
sum rule[9] $$ s_{\beta}^2[(\hat{S}_i^t)^2+(\hat{P}^t_i)^2]+c_{\beta}^2
[(\hat{S}_i^b)^2+(\hat{P}^b_i)^2] \eqno(19) $$ is immediate.

The couplings ZZ-Higgs are $g_{ZZh_i}=\frac{gm_Z}{c_w}C_i$ with $$
C_1=-s_{\psi}s_{2\beta}\ \ \ \ \ \ C_2=c_{\psi}s_{2\beta}\ \ \ \ \ \ 
C_3=c_{2\beta} \eqno(20) $$ For the couplings Z-Higgs-Higgs we have 
$g_{Zh_ih_j}=\frac{g}{2c_w }C_{ij}$ with $$
C_{12}=c_{2\beta}\ \ \ \ \ \ C_{23}=-s_{\psi}s_{2\beta}\ \ \ \ \ \ 
C_{31}=c_{\psi}s_{2\beta} \eqno(21) $$ antisymmetric in the indices. 
Clearly $$ C_i=C_{jk} \eqno(22) $$ for $(i,j,k)$ being any permutation 
of (1,2,3). The sum rule $\sum_k{C_k^2}=1$ becomes $$ 
C_i^2+C_j^2+C_{ij}^2=1 \eqno(23) $$ $(i\ne{j};i,j=1,2,3)$ which requires 
at least one of the ZZ$h_i$, ZZ$h_j$ and Z$h_ih_j$ couplings to be 
substantial in size. At last, one can very easily check also the 
following sum rules $$ 
2\hat{S}^t_is_{\beta}^2+C_i=2\hat{S}_i^bc_{\beta}^2+C_i=0\ \ \ \ \ \ 
(i=1,2) $$ $$ 
2\hat{S}^t_3s_{\beta}^2+C_3=-(2\hat{S}^b_3c_{\beta}^2+C_3)=1 $$
in agreement with the general sum rules[9] $$
(\hat{S}_i^t)^2+(\hat{P}_i^t)^2=\frac{1}{t_{\beta}^2}[1+C_i(2\hat{S}^b_i+\frac
{C_i}{c_{\beta}^2})] $$ $$ 
(\hat{S}_i^b)^2+(\hat{P}_i^b)^2=t_{\beta}^2[1+C_i(2\hat{S}_i^t+\frac{C_i}
{s_{\beta}^2})] \eqno(24) $$ (i=1,2{\em and} 3).

In ref.[9], it is well shown that the set of these sum rules makes it 
apparent that the Higgs finding strategy at $e^+e^-$ colliders should 
include the Yukawa processes with Higgs radiation off top and bottom 
quarks (the resort to heavy flavour sector is obvious) together with 
Higgs-strahlung and Higgs pair production.

\section{\bf Higgs boson production in $e^+e^-$ colliders}

We consider now the production of the neutral Higgs boson $h_i$ in 
association with a fermion pair in $e^+e^-$ collisions $$ 
e^+e^-\to{f\overline{f}h_i} $$
It can proceed via (1) bremsstrahlung off the Z boson 
$e^+e^-\to{Z^*h_i}\to{f\overline{f}h_i}$, (2) Higgs pair production 
$e^+e^-\to{h_j^*h_i}\to{f\overline{f}h_i}$, 
$e^+e^-\to{h_k^*h_i}\to{f\overline{f}h_i}$ ($(i,j,k)$ any permutation 
(1,2,3)) and (3) Yukawa processes with Higgs radiation off a fermion 
line $e^+e^-\to{f\overline{f}}^*\to{f\overline{f}h_i}$, 
$e^+e^-\to{f^*\overline{f}}\to{f\overline{f}h_i}$. (On the importance of 
the Yukawa processes in a CP conserving 2HDM see [13], [14].) 

The total cross section $\sigma(e^+e^-\to{f\overline{f}h_i})\equiv
\sigma(f\overline{f}h_i)$ of our process can be 
written as follows. Let $N_c$ denote the number of colors, 
$\sigma_0=4\pi\alpha^2/s$ the standard normalization cross section, 
where $\sqrt{s}$ is the total c.m. energy, $q_f$ the electric charge, 
$v_f$ and $a_f$ the vector and axial Z charges of the fermion $f$ $$
v_f=\frac{2I_{L,f}-4q_fs_W^2}{4s_Wc_W}\ \ \ \ \ \ \ \ 
a_f=\frac{2I_{L,f}}{4s_Wc_W} \eqno(25) $$ with $I_{L,f}=\pm\frac{1}{2}$ 
being the weak isospin of the left-handed fermions. We have
$$ 
\sigma(f\overline{f}h_i)=N_c\frac{\sigma_0}{(4\pi)^2}\int_{4f}
^{(1-\sqrt{h_i})^2}{dy}\{[q_e^2q_f^2+2q_fv_fI(z)+(v_f^2+a_f^2)Q(z)]H_0
^{(1)}(y)+ $$ $$
+Q(z) [a_f^2(H_0^{(2)}(y)+H_1^{(1)}(y)+H_2^{(1)}(y))+(v_f^2+a_f^2)H_1^{(2)}(y)
+H_3(y)]+ $$ $$ +[q_fv_fI(z)+(v_f^2+a_f^2)Q(z)]H_2^{(2)}(y)\} \eqno(26) $$ 
In eq.(26) $$ 
I(z)=\frac{q_ev_e(1-z)}{(1-z)^2+\gamma_zz}\ \ \ \ \ \ \ 
Q(z)=\frac{v_e^2+a_e^2}{(1-z)^2+\gamma_zz} $$
where $z=\frac{m_Z^2}{s}$, $\gamma_z=\frac{\Gamma_Z^2}{s}$ are 
the reduced mass and 
width of the Z boson. Analogously,  
$f=\frac{m_f^2}{s}$, $h_i=\frac{m(h_i)^2}{s}$ are the reduced masses of
the fermion and the $h_i$ neutral Higgs boson, and in the following we 
shall use $\gamma_i=\frac{\Gamma(h_i)^2}{s}$ for the reduced width of $h_i$.

In order to specify the $H_m^{(n)}(y)$ functions, we introduce the 
triangular function $$
\lambda_i\equiv\lambda(1,h_i,y)=(1+h_i-y)^2-4h_i \eqno(27) $$   
and $$ \Delta_i\equiv\left[\frac{y-4f}{y}\lambda_i\right]^{1/2}\ \ \ \ \ \ \ 
B_i\equiv\ln\frac{1+h_i-y+\Delta_i}{1+h_i-y-\Delta_i} \eqno(28) $$
and the combinations of the scalar $S_i^f$ and pseudoscalar $P_i^f$ 
couplings $$
\alpha_i^{f(\pm)}=(S_i^f)^2\pm{(P_i^f)^2} \eqno(29) $$ We have $$ 
H_0^{(1)}(y)=2\alpha_i^{f(+)}\left[\frac{(2f-h_i)(1+2f)y}{f\lambda_i+h_iy}
\Delta_i+\frac{(1+2f-y)^2+(2f-h_i)^2+4f}{1+h_i-y}B_i\right]+ $$ $$ 
+4f\alpha_i^{f(-)}\left[\frac{(1+2f)y}{f\lambda_i+h_iy}\Delta_i+2
\frac{2+2f-y}{1+h_i-y}B_i\right] \eqno(30a) $$ 
$$ 
H_0^{(2)}(y)=2\alpha_i^{f(+)}\left[\left(y-2-\frac{6f(2f-h_i)y}{f\lambda_i
+h_iy}\right)\Delta_i-2\frac{f\lambda_i+2(3f-h_i)(1+2f-y)+6fh_i}{1+h_i-y}
B_i\right]+ $$ $$
-24f\alpha_i^{f(-)}\left[\frac{fy}{f\lambda_i+h_iy}\Delta_i+\frac
{1+2f-y}{1+h_i-y}B_i\right] \eqno(30b)  $$ 
$$ H_1^{(1)}(y)=\frac{g_{ZZh_i.}^2}{(y-z)^2+\gamma_zz}2f[-12z+(\frac{y}{z}-2)
\lambda_i]\Delta_i \eqno(30c) $$ 
$$ H_1^{(2)}(y)=\frac{g_{ZZh_i.}^2}{(y-z)^2+\gamma_zz}4z(y+2f)(1+
\frac{\lambda_i}{12y})\Delta_i \eqno(30d) $$ $$ 
H_2^{(1)}(y)=S_i^fg_{ZZh_i.}\frac{y-z}{(y-z)^2+\gamma_zz}4\sqrt{\frac{f}{z}}
\{[(1+h_i-y)y-6z]\Delta_i+ $$ $$ -2[3z(4f-h_i)+f\lambda_i+h_iy]B_i\} 
\eqno(30e) $$ 
$$H_2^{(2)}(y)=S_i^fg_{ZZh_i.}\frac{y-z}{(y-z)^2+\gamma_zz}8\sqrt{fz}[2\Delta_i
+(1-2h_i+4f+y)B_i] \eqno(30f) $$ $$ 
H_3(y)=\left\{\frac{[S_j^fC_{ij}(y-h_k)+(j\leftrightarrow{k})]^2(y-4f)
+[P^f_jC_{ij}(y-h_k)+(j\leftrightarrow{k})]^2y}{8s_W^2c_W^2[(y-h_j)^2+
\gamma_jh_j][(y-h_k)^2+
\gamma_kh_k]}+\right. $$ $$   
+\left.\frac{a_fg_{ZZh_i.}}{s_Wc_W}\sqrt{\frac{f}{z}}\left[
\frac{P_j^fC_{ij}(y-h_j)}
{(y-h_j)^2+\gamma_jh_j}+(j\to{k})\right
]\right\}\lambda_i\Delta_i+ \eqno(30g) $$ $$+\frac{a_f}{s_Wc_W}\left[
\frac{(S_i^fP_j^f+S_j^fP_i^f)C_{ij}(y-h_j)}{(y-h_j)^2+\gamma_jh_j}+(j\to{k})
\right][y(1+h_i-y)\Delta_i-2(f\lambda_i+h_iy)B_i]. $$
$(i,j,k)$ being any permutation of (1,2,3).

The functions $H_0^{(1)}(y),H_0^{(2)}(y)$ describe the Yukawa processes 
with $h_i$ neutral Higgs radiation off the fermions. The functions 
$H_1^{(1)}(y),H_1^{(2)}(y)$ describe the Higgs-strahlung off the Z 
boson. The functions $H_2^{(1)}(y),H_2^{(2)}(y)$ describe the 
interference between the above mentioned processes. The function 
$H_3(y)$ describe the Higgs pair production (2) and its interference 
with the other two mechanisms of $h_i$ production.

This result is in right agreement with ref.[9]. A part a $1/4\pi$
common factor, already taken into account, the $H_m^{(n)}$ 
functions are integrated forms of the functions $F_n,G_n$ or their 
combinations. 

\section{\bf Numerical results}

We can now proceed to comment our results with some examples.

As we argued above, the violation of CP in the 2HDM taken into account, 
 plays a r\^{o}le not only in determining the couplings 
of the neutral Higgs bosons but also in adjusting their mass spectrum.
For growing $\psi=\frac{\theta}{2}$, the tendency of the masses 
towards the lightest one, makes more interesting the 
$\sum_{1=k}^3\sigma(f\overline{f}h_k)$ (with $f=t,b$)  rather than the 
single $\sigma(f\overline{f}h_k)$ cross sections. 

In the numerical examples, we have taken $m_b=5$Gev, $m_t=175$Gev, 
$\gamma_k=0.02m(h_k)$. Moreover, whenever $m(h_1)$ is given, then 
 $m(h_1)=100$Gev.

The fractional contributions of the single $\sigma(t\overline{t}h_k)$ 
cross sections to their sum, is shown in Fig.1. Their dependence on 
$\psi$, for given $\tan\beta$, is obvious owing to the influence of $\psi$
both on the production thresholds and on the coupling constants. In this 
respect, one can compare the panels (a) and (b) and the panels (c) and 
(d) of Fig.1

 In the same figure, the comparison of the panels (a) and 
(c) and of the panels (b) and (d) exhibits the dependence on 
$\tan\beta$ of the fractional contributions 
$\sigma(t\overline{t}h_k)$, for given $\psi$. Clearly, the increase
of $\tan\beta$ grows the fractional 
contribution of $h_3$ (whose coupling to quarks do not depend on 
$\tan\beta$ in our model)  with respect to those 
of the other two neutral Higgs, owing to the break down of their 
coupling to up-type quarks for large $\tan\beta$.
 
As regards $\sum_k\sigma(b\overline{b}h_k)$, an opposite remark
holds, as it is apparent by 
comparison of the panels (a) and (c) and the panels (b) and (d) of Fig.2.
Now, for given $\psi$, the rise of $\tan\beta$ reduces the fractional 
contribution of $h_3$, owing to the independence on $\tan\beta$ of its 
coupling to quarks, while the couplings of  $h_1$ and $h_2$ to down-type 
quarks are enhanced for large $\tan\beta$. 

The dependence on $\psi$ of the fractional contributions 
$\sigma(b\overline{b}h_k)$ to their sum, for given $\tan\beta$, 
is particularly evident by 
comparison of the panels (a) and (b) and of the panels (c) and (d) of 
Fig.2.  For rising $\psi$, the production thresholds of 
$b\overline{b}h_2$ and $b\overline{b}h_3$ move backwards and strikingly 
modifie the profile of $\sum_k\sigma(b\overline{b}h_k)$ versus the 
total c.m. energy.    

As an example of these effects, we note that the hunting of the neutral 
Higgs bosons at $\sqrt{s}=500$Gev, $\tan\beta=10$, and $\psi=27^o$, is 
practically the hunting of $h_3$, the most massive(!) of the neutral 
Higgs bosons in our model.

The Fig.3 shows the behaviour of $\sum_k\sigma(t\overline{t}h_k)$ versus 
$\sqrt{s}$ for different choices of $\psi$, for given $\tan\beta$. Its 
growing with $\psi$ is clearly shown. 

The Fig.4 shows the behaviour of $\sum_k\sigma(b\overline{b}h_k)$ versus 
$\sqrt{s}$ for given $\tan\beta$ and various values of $\psi=\theta/2$, 
which parametrizes in some way the CP breaking. In this figure the curve 
for the CP conserving case ($\psi=0$) is invisible in the chosen scale.
For growing $\psi$, 
the lowering of the production thresholds and the growth of 
$\sum_k\sigma(b\overline{b}h_k)$ is also well evident.

In Fig.5 one can see the dependence of 
$\sum_k\sigma(t\overline{t}h_k)$ on the parameter $\psi$ of CP breaking, 
at given $\tan\beta$, for different choices of the c.m. total energy 
$\sqrt{s}$. The common feature of these curves (a part that they are 
 symmetric in $\psi$) is their remarkable increase with $\psi(>0)$, at 
least if $\psi$ is not too small. Similar remarks hold  
 in Fig.6 where is shown the behaviour of $\sum_k\sigma(b\overline{b}h_k)$ 
versus $\psi$, at given $\tan\beta$ and for various $\sqrt{s}$. In this 
case, it is well evident that the width of the plateau around 
$\psi=0$ (where the gain of 
$\sum_k\sigma(b\overline{b}h_k)$ with respect to the CP conserving case 
$\psi=0$ is unsizeable) decreases as $\sqrt{s}$ increases.

In the Figs.7 and 8, are reported the behaviours of 
$\sum_k\sigma(t\overline{t}h_k)$ and $\sum_k\sigma(b\overline{b}h_k)$ 
versus $m(h_1)$, at given $\sqrt{s}$ and at some values of $tan\beta$ 
and $\psi$. At given $\tan\beta$, the minimal values of 
$\sum_k\sigma(t\overline{t}h_k)$($\sum_k\sigma(b\overline{b}h_k)$) are 
realized in the CP conserving case $\psi=0$. In effect, in this case, 
only $\sigma(t\overline{t}h_1)$($\sigma(b\overline{b}h_1)$) contributes, 
because the other two neutral Higgs are enormously massive. In any case, 
the maximal values are achieved for "maximal" CP violation, that is for 
$\psi=\frac{\pi}{4}(\theta=\frac{\pi}{2})$. Independently on the mass of 
the lightest Higgs, this confirms what we have already seen for 
$m(h_1)=100Gev$. Also the occurrence that the smaller is $\tan\beta$
 (at given $\psi$) and the 
larger is the $\sum_k\sigma(t\overline{t}h_k)$ (the smaller is the 
$\sum_k\sigma(b\overline{b}h_k)$), is 
consistent with the previous considerations concerning the dependence on 
$\tan\beta$ of the couplings of the $h_1$ and $h_2$ bosons to up-type 
and down-type quarks.

But we can well note in Fig.8 that the maximal 
$\sum_k\sigma(b\overline{b}h_k)$ at $\tan\beta=10$ can be larger than 
the minimal one at $\tan\beta=100$. In other terms, a small $\tan\beta$ 
and a large $\psi$ can yield a cumulative cross section 
$\sum_k\sigma(b\overline{b}h_k)$ larger than the one with a larger 
$\tan\beta$ but a smaller $\psi$. This can be observed also for 
$\Sigma_t=\sum_k\sigma(t\overline{t}h_k)$ in the panel (a) of Fig.9, 
where the contour lines for $\Sigma_t$ are reported in the plane 
$(\tan\beta,\psi)$, at given $m(h_1)=100Gev$ and $\sqrt{s}=600Gev.$ 
E.g., for $\tan\beta=0.1$ and $\psi=30^0$ one has a $\Sigma_t$ larger 
than for $\tan\beta=0.07$ and $\psi=10^o$: the growing of $tan\beta$ 
yields effects opposite to those due to the growing of $\psi$.

The panels (b) and (c) of Fig.9 show the contour lines for the cumulative 
cross section $\Sigma_t$ for two different choices of $\tan\beta$, at 
given $\sqrt{s}=600Gev$, in the plane $(m(h_1),\psi)$. Also here, a 
small $m(h_1)$ is not a guarantee of a great $\Sigma_t$: that depends on 
$\psi$. E.g., $m(h_1)=60Gev,\psi=5^o$ has a $\Sigma_t$ smaller than 
the one for $m(h_1)=150Gev,\psi=45^o$ (at 
$\tan\beta=0.1,\sqrt{s}=600Gev$). Analogous considerations can be made 
concerning $\Sigma_b=\sum_k\sigma(b\overline{b}h_k)$, as it is shown in 
the panel (d) of Fig.9.

In our 2HDM, at given $m(h_1)$, to change $\psi$ is equivalent to change 
$m(h_2)$ (and $m(h_3)$). Therefore the former figures can be turned in 
the plane $(m(h_1),m(h_2))$ (or $(m(h_1),m(h_3))$). They are shown in 
Fig.10. In the panel (a) are reported the contour lines for $\Sigma_t$, 
 at $\tan\beta=0.1$ and c.m. energy $\sqrt{s}=600Gev$, in the plane 
$(m(h_1),m(h_2))$ (solid lines) and $(m(h_1),m(h_3))$ (dashed lines). 
Analogously in the panel (b) one can see the contour lines for
 $\Sigma_b$ (with $\tan\beta=10$ and $\sqrt{s}=400Gev$).
\vspace{0.7cm}

We can conclude that the spontaneous CP violation in our 2HDM not only 
do not jeopardize but, on the contrary, increases our ability to find 
light neutral Higgs bosons. Furthermore, to find a neutral Higgs boson 
do not necessarily mean to have found the lightest one: it can happen 
that the largest contribution to this hunting is due to the heaviest 
neutral Higgs boson. 

\bibliographystyle{plain}

\vfill\eject
\bigskip
{\Large\bf Figure captions}

\begin{itemize} 
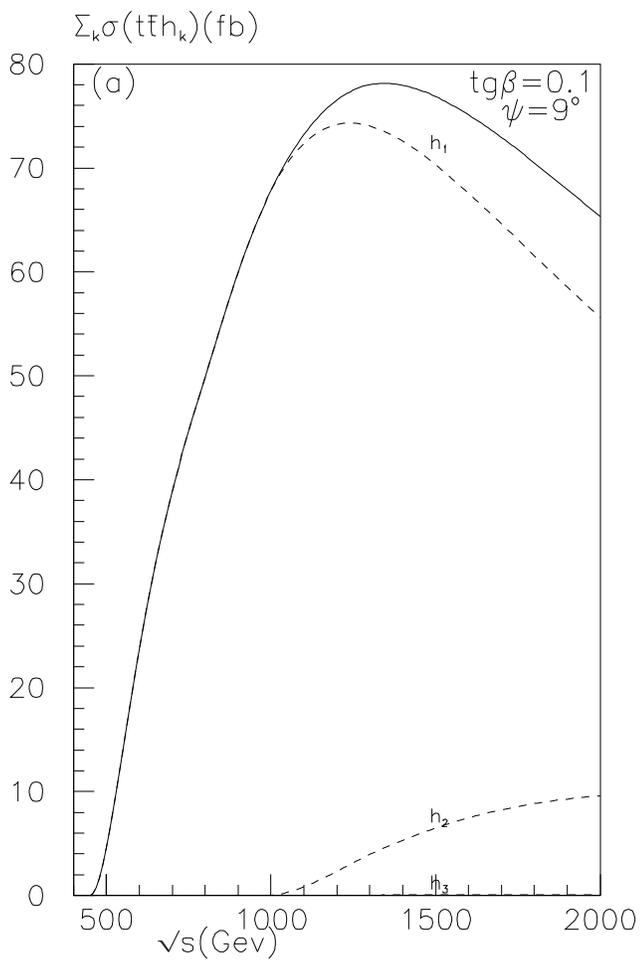
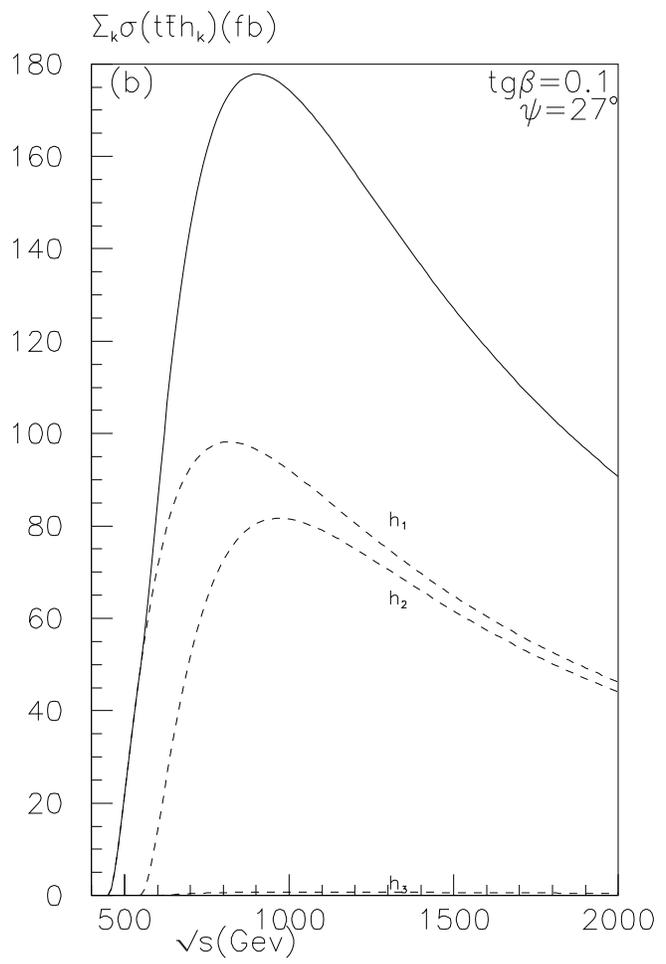
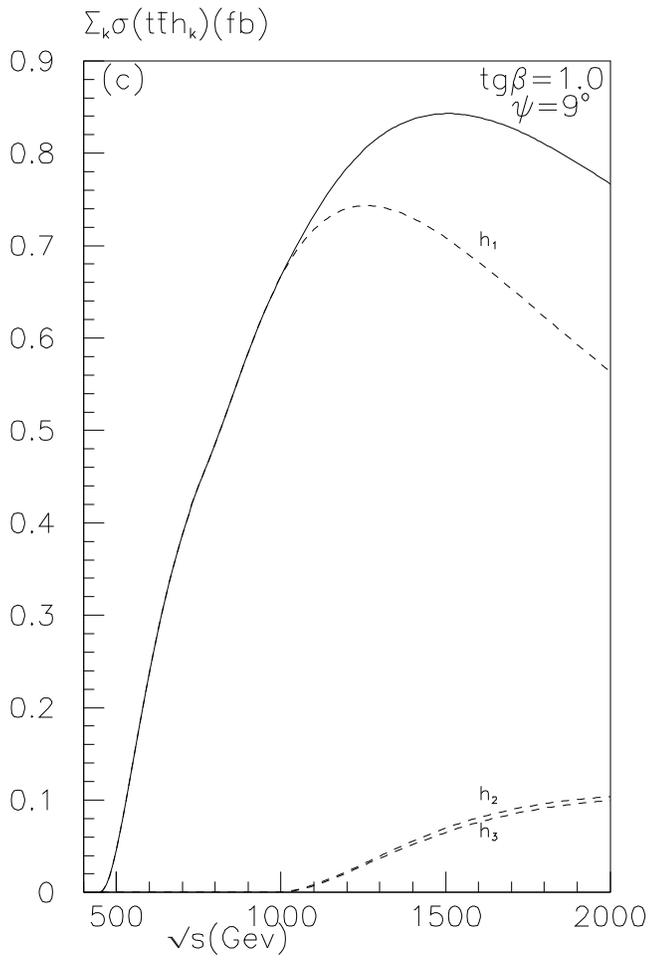
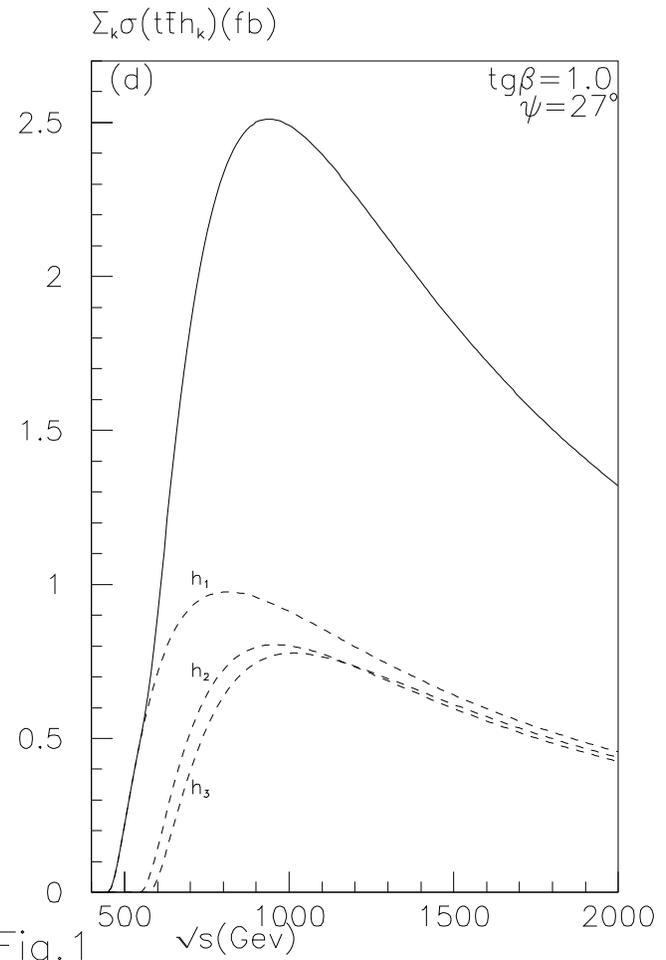
\item[{\bf Fig.1}] The cumulative 
$\sum_{1=k}^3\sigma(e^+e^-\to{t\overline{t}h_k})\equiv\sum_k
\sigma(t\overline{t}h_k)$ cross section and its contributions 
$\sigma(t\overline{t}h_i)$, $i=1,2,3$, versus the c.m. energy 
$\sqrt{s}$, for $m(h_1)=100Gev$ and {\bf (a)} $\tan\beta=0.1, \psi=9^o$; 
{\bf (b)} $\tan\beta=0.1, \psi=27^o$; {\bf (c)} $\tan\beta=1.0, 
\psi=9^o$; {\bf (d)} $\tan\beta=1.0, \psi=27^o$. 
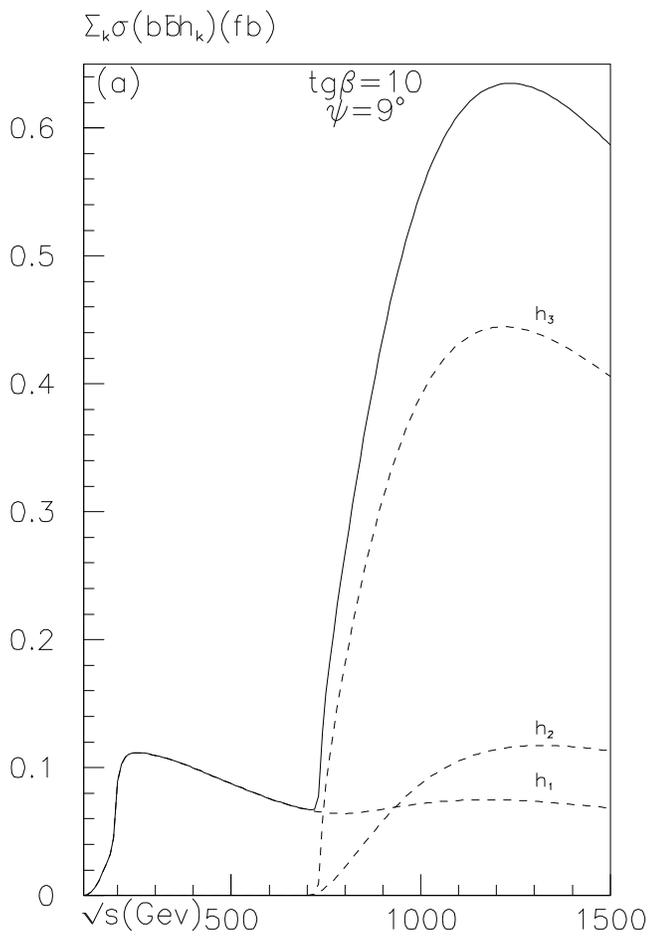
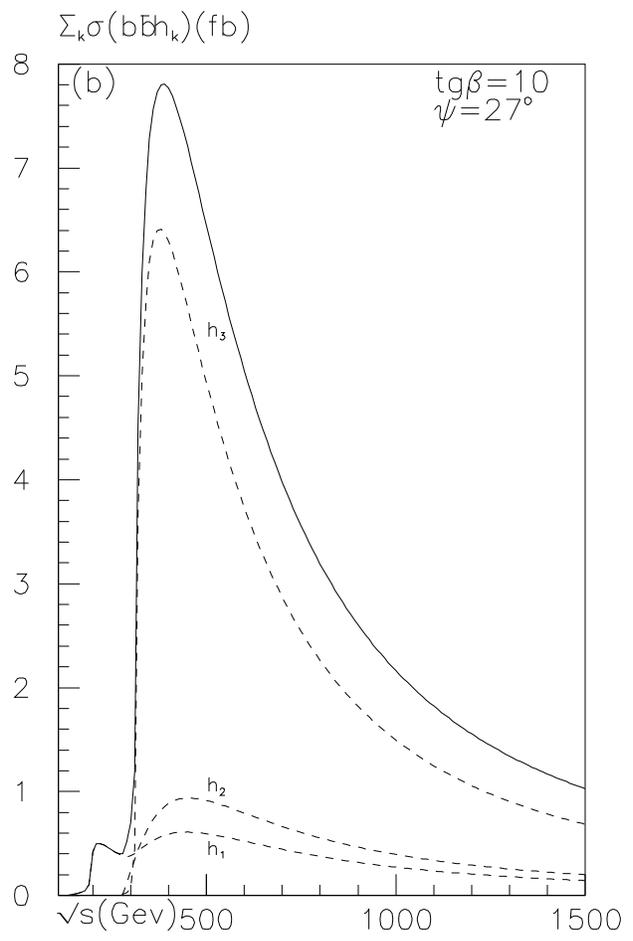
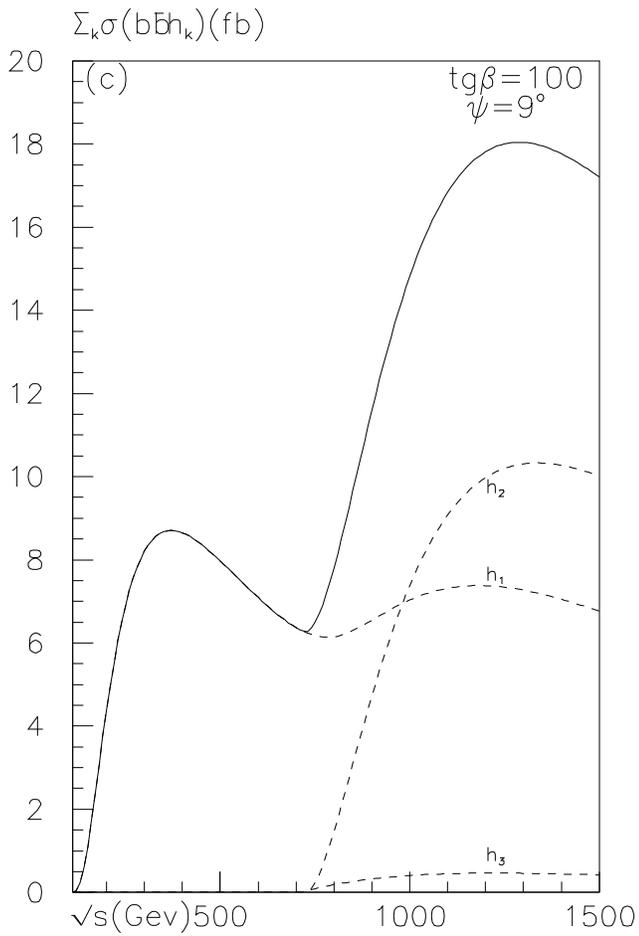
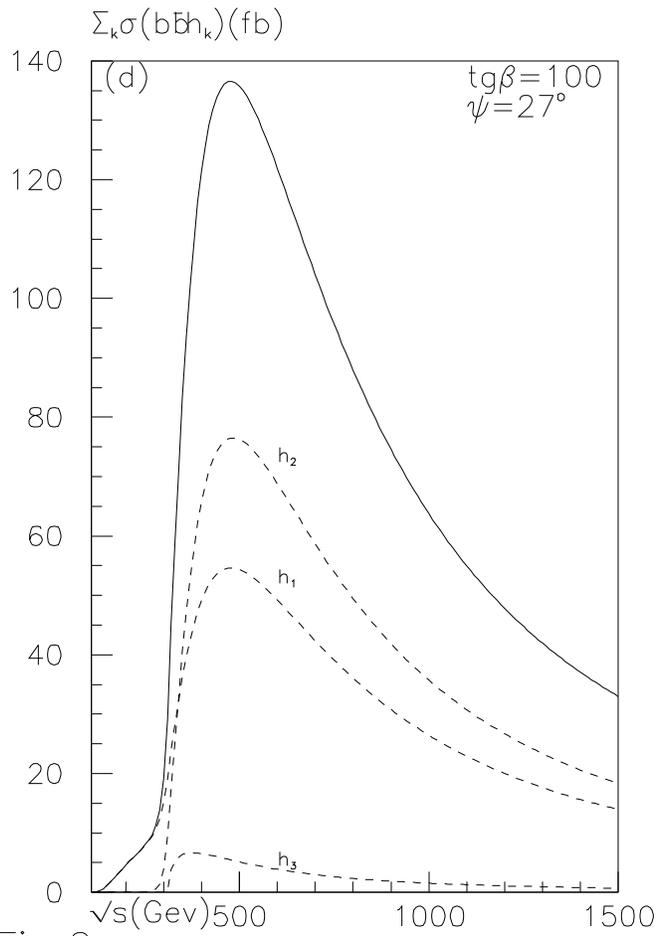
\item[{\bf Fig.2}] The cumulative 
$\sum_{1=k}^3\sigma(e^+e^-\to{b}\overline{b}h_k)\equiv\sum_k\sigma(b\overline
{b}h_k)$ cross section and its contributions $\sigma(b\overline{b}h_i)$, 
$i=1,2,3,$ versus the c.m. energy $\sqrt{s}$, for $m(h_1)=100Gev$ and 
{\bf (a)} $\tan\beta=10, \psi=9^o$; {\bf (b)} $\tan\beta=10, \psi=27^o$; 
{\bf (c)} $\tan\beta=100, \psi=9^o$; {\bf (d)} $\tan\beta=100, 
\psi=27^o$.
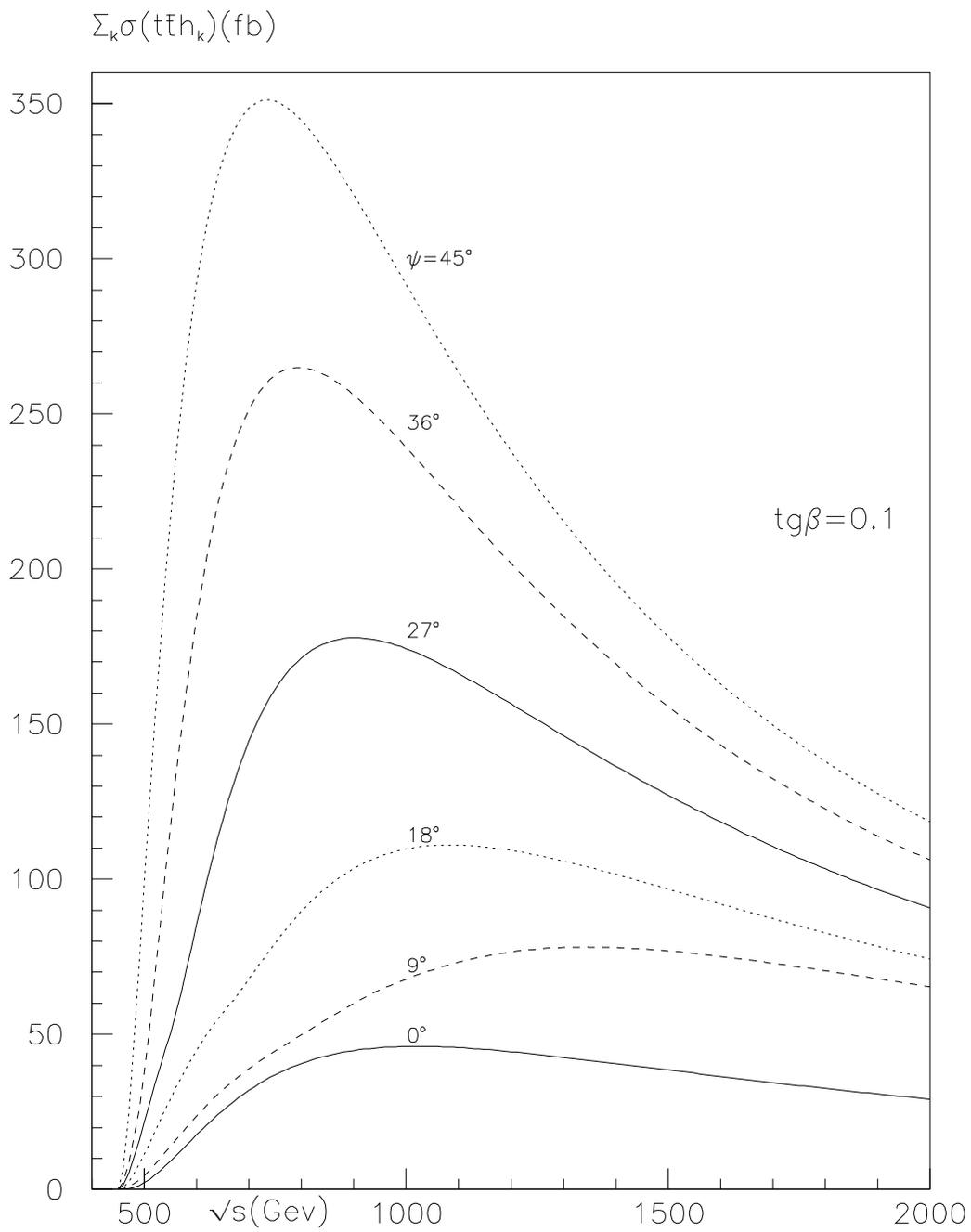
\item[{\bf Fig.3}] The $\sum_k\sigma(t\overline{t}h_k)$ cross section 
versus $\sqrt{s}$, for $m(h_1)=100Gev$, $\tan\beta=0.1$ and for various 
values of $\psi$, from $\psi=0^o$ (CP conserving case) to $\psi=45^o$ 
("maximal" CP breaking).
\item[{\bf Fig.4}] The $\sum_k\sigma(b\overline{b}h_k)$ cross section 
versus $\sqrt{s}$, for $m(h_1)=100Gev$, $\tan\beta=10$ and for various 
values of $\psi$. The curve corresponding to the CP conserving case 
$\psi=0^o$ is undrawn because it is too small in the chosen scale.
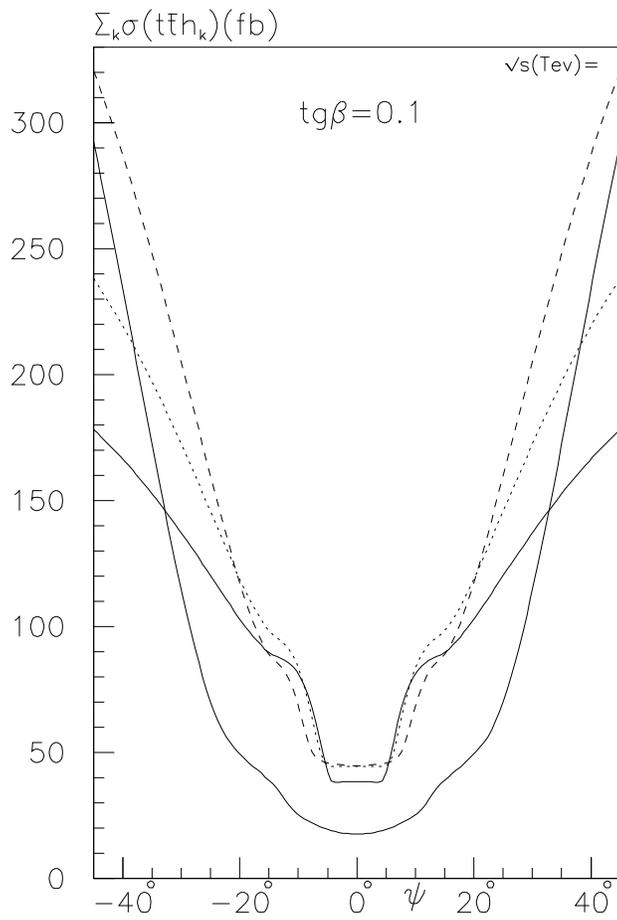
\item[{\bf Fig.5}] The dependence on $\psi$ of the 
$\sum_k\sigma(t\overline{t}h_k)$ cross section, for $m(h_1)=100Gev$, 
$\tan\beta=0.1$ and for various values of the c.m. energy $\sqrt{s}$.
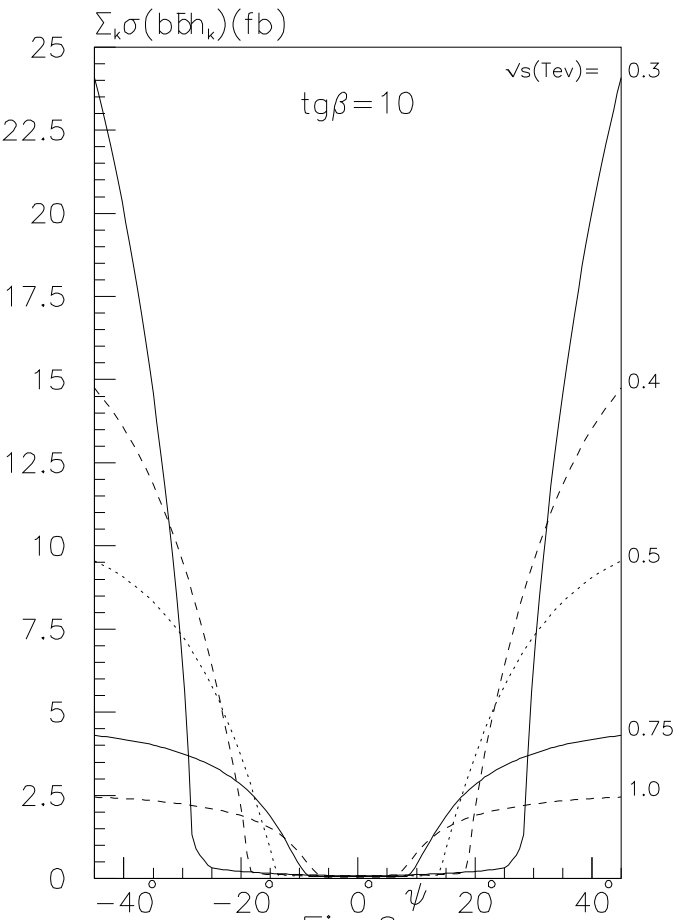
\item[{\bf Fig.6}] The $\sum_k\sigma(b\overline{b}h_k)$ cross section 
versus $\psi$, for $m(h_1)=100Gev$, $\tan\beta=10$ and for various 
values of $\sqrt{s}$. 
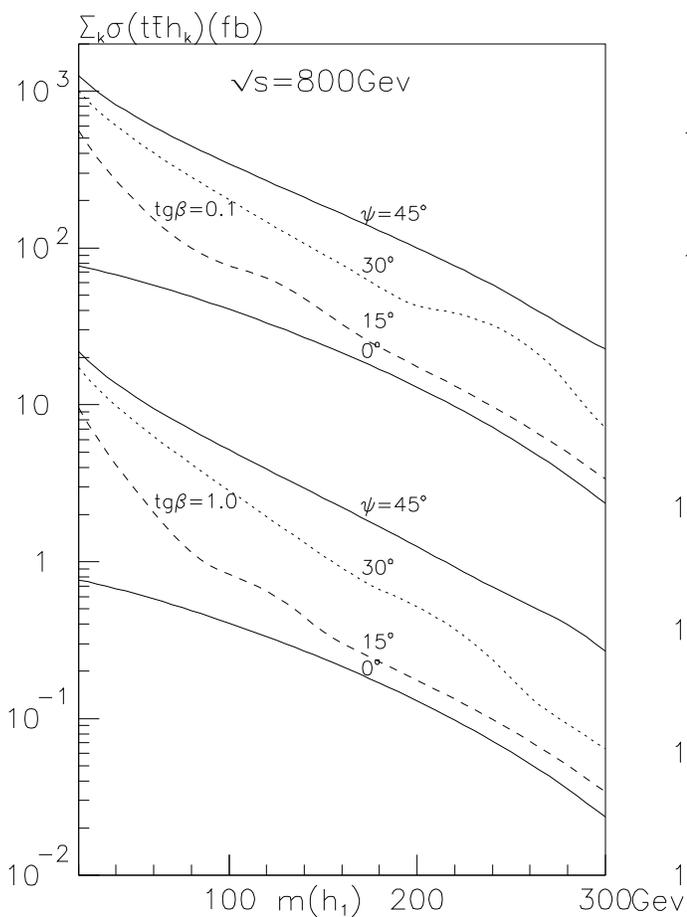
\item[{\bf Fig.7}] The cumulative $\sum_k\sigma(t\overline{t}h_k)$ cross 
section versus the mass of the lightest neutral Higgs boson $m(h_1)$, at 
given $\sqrt{s}=800Gev$, for $\tan\beta=0.1$ and various values of 
$\psi$, and for $\tan\beta=1.0$ and various values of $\psi$.
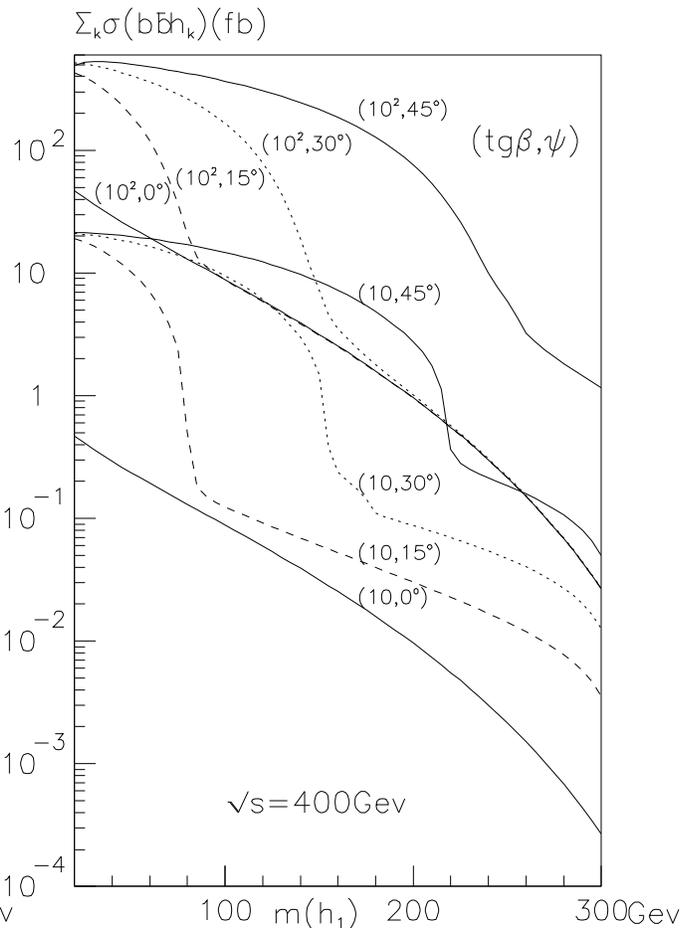
\item[{\bf Fig.8}] the $\sum_k\sigma(b\overline{b}h_k)$ cross section 
versus $m(h_1)$, at $\sqrt{s}=400Gev$. There are two sets of lines: for 
$\tan\beta=10$ and various values of $\psi$, and for $\tan\beta=100$ and 
various values of $\psi$.
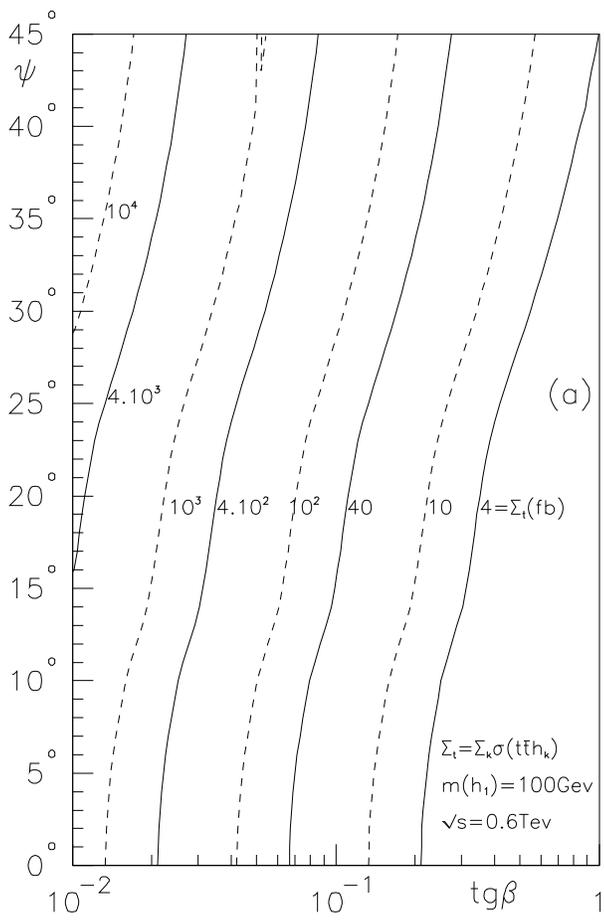
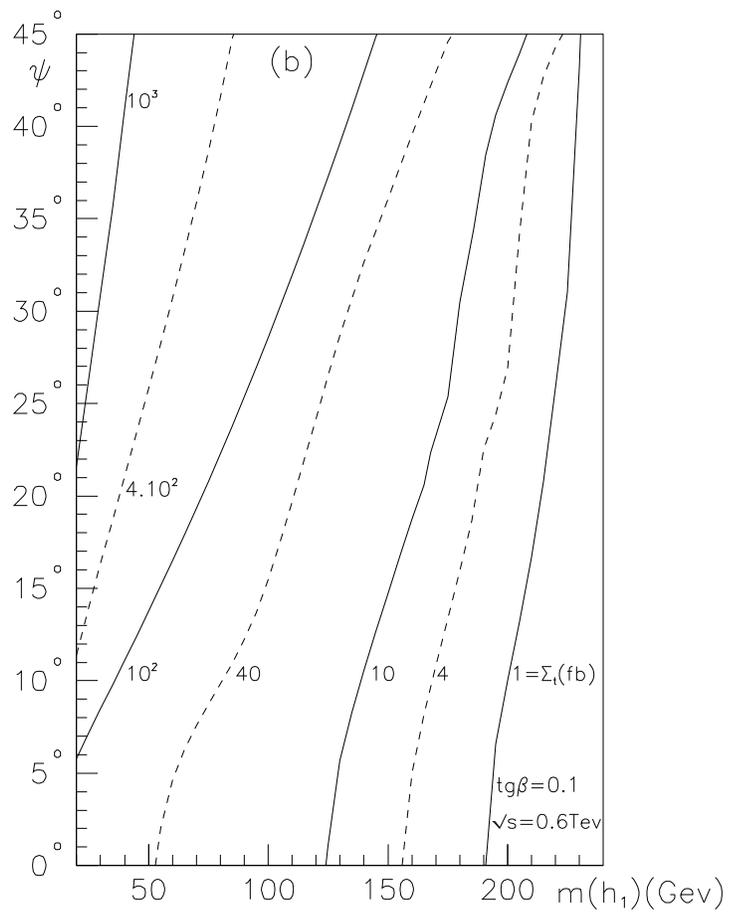
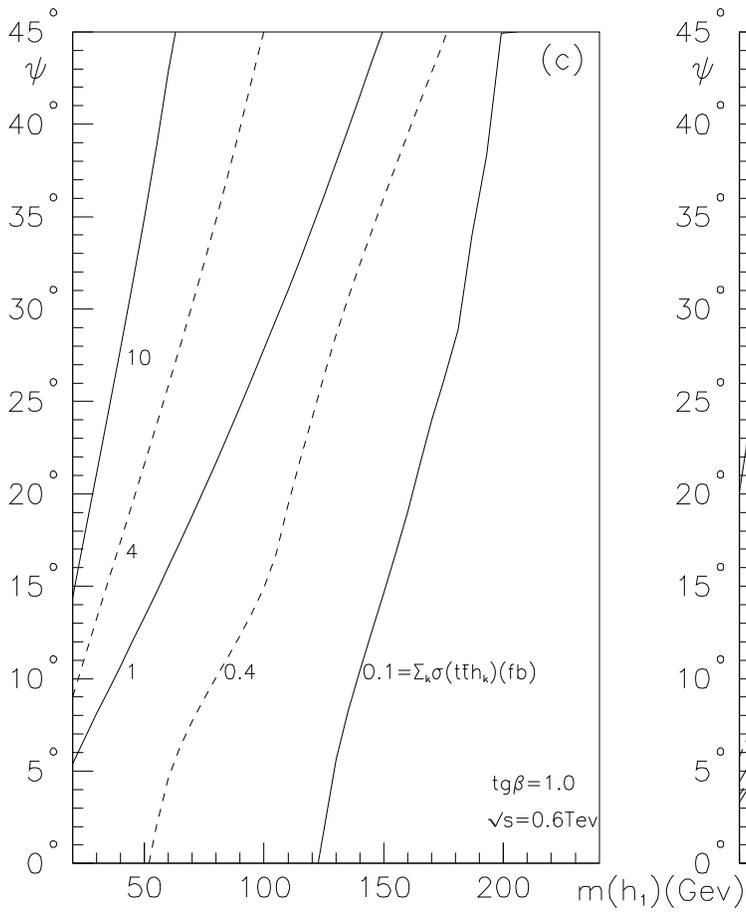
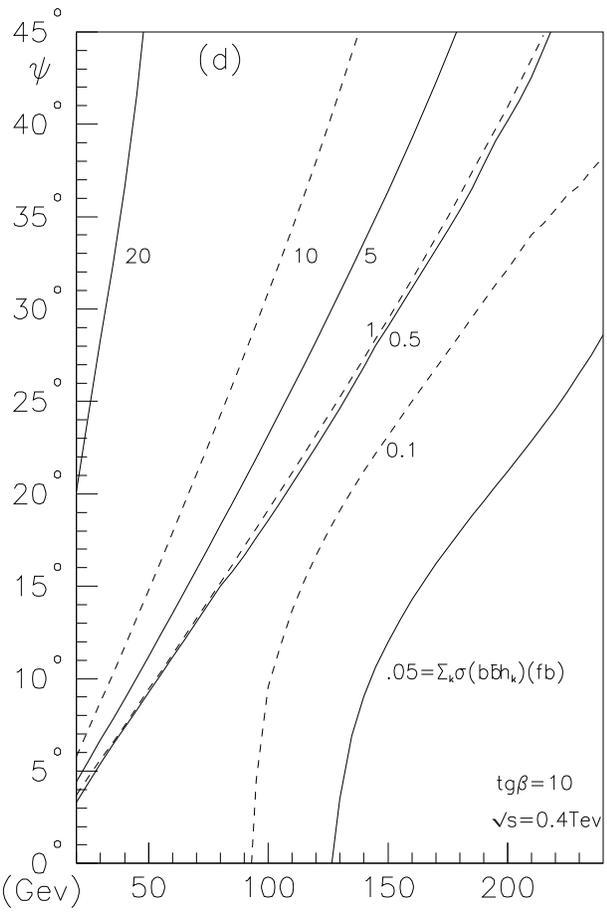
\item[{\bf Fig.9}] {\bf (a)} Contour lines for 
$\sum_k\sigma(t\overline{t}h_k)\equiv\Sigma_t$ in the plane 
$(\tan\beta,\psi)$ with $m(h_1)=100Gev$ and $\sqrt{s}=600Gev$; {\bf (b)} 
contour lines for $\Sigma_t$ in the plane $(m(h_1),\psi)$, for 
$\tan\beta=0.1$ and $\sqrt{s}=600Gev$; {\bf (c)} same of (b) but for 
$\tan\beta=1.0$; {\bf (d)} contour lines for
$\sum_k\sigma(b\overline{b}h_k)\equiv\Sigma_b$ in the plane 
$(m(h_1),\psi)$, for $\tan\beta=10$ and $\sqrt{s}=400Gev$.
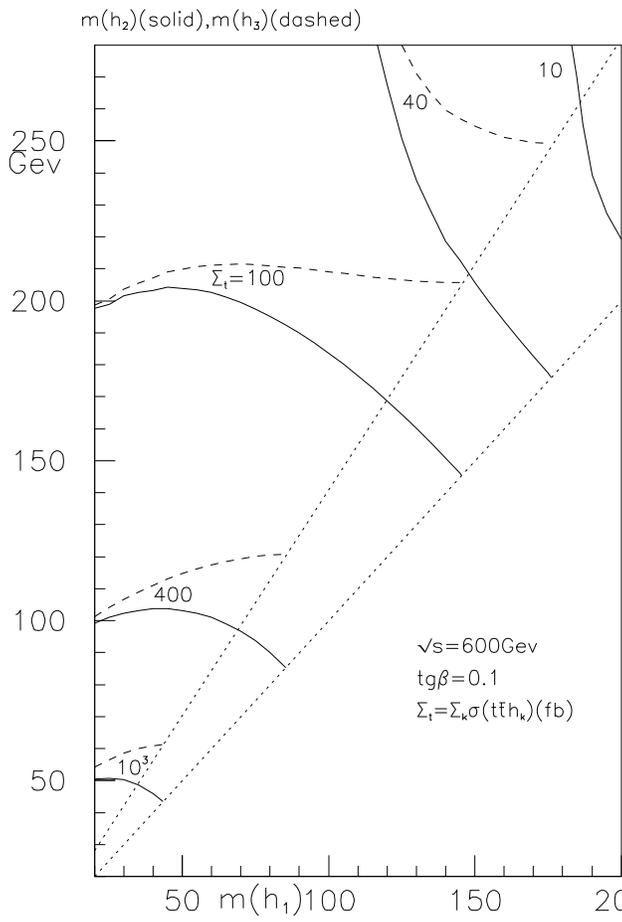
\item[{\bf Fig.10}] {\bf (a)} Contour lines for 
$\Sigma_t=\sum_k(t\overline{t}h_k)$ in the plane $(m(h_1),m(h_2))$ 
(solid lines) or in the plane $(m(h_1),m(h_3))$ (dashed lines). Here 
$\tan\beta=0.1$ and $\sqrt{s}=600Gev$; {\bf (b)} same of (a) but for 
$\Sigma_b=\sum_k\sigma(b\overline{b}h_k)$, with $\tan\beta=10$ and 
$\sqrt{s}=400Gev$.
\end{itemize} 
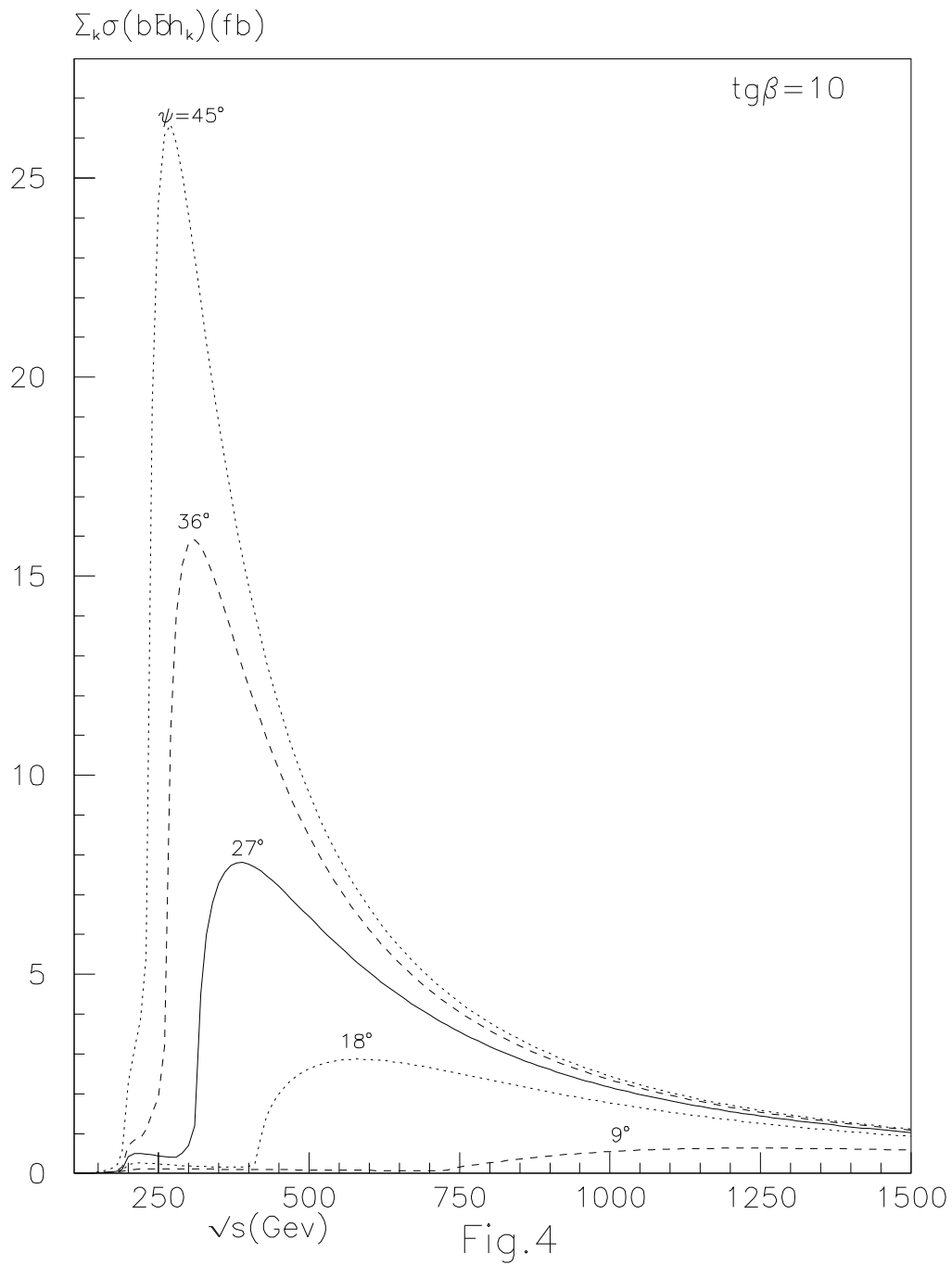
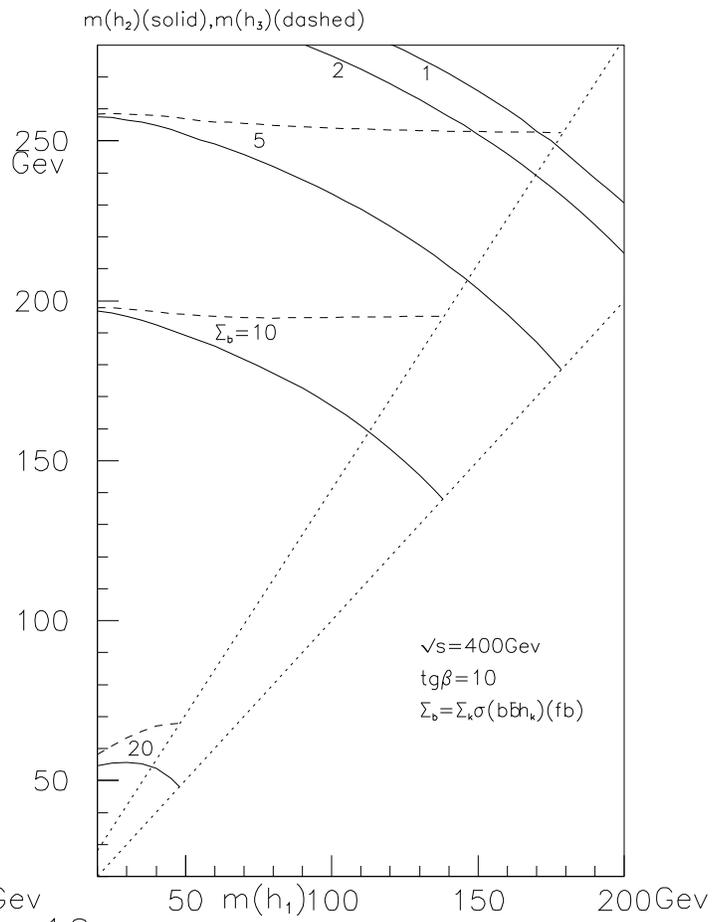
\end{document}